\documentclass[aps,showpacs,preprintnumbers]{revtex4}
\usepackage{bm}

\input{tcilatex}

\begin{document}

\title{Revisiting Bohr's quantization hypothesis for the atomic orbitals}
\author{J.H.O.Sales}
\affiliation{$^{1}$Funda\c{c}\~{a}o de Ensino e Pesquisa de Itajub\'{a} , Av. Dr. Antonio
Braga Filho, CEP 37501-002, Itajub\'{a}, MG.\\
$^{2}$Instituto de Ci\^{e}ncias, Universidade Federal de Itajub\'{a}, CEP
37500-000, Itajub\'{a}, MG, Brazil.}
\author{A.T. Suzuki}
\affiliation{Department of Physics, North Carolina State University, Raleigh, NC
27695-8202.}
\author{D.S. Bonaf\'{e}}
\affiliation{Escola Estadual Major Jo\~{a}o Pereira, Av. Paulo Chiaradia, Cep: 37500-028,
Itajub\'{a}, MG, Brazil.}
\date{\today }

\begin{abstract}
We deduce the quantization of the atomic orbit for the hydrogen's atom model
proposed by Bohr without using his hypothesis of angular momentum
quantization. We show that his hypothesis can be deduced from and is a
consequence of the Planck's energy quantization.
\end{abstract}

\maketitle

%\preprint{APS/123-QED}






\section{Introduction}

\noindent The existence of an atomic nucleus was confirmed in 1911 by E.
Rutherford in his classic scattering experiment \cite{ruth}. Before that
time, it was believed that an atom was something like a positively charged
dough and negatively charged raisins scattered here and there within the
dough, or an ice-cream with chocolate chip flakes in it, where the ice-cream
would be the positive charge (protons) and the chocolate chip flakes the
negative charge (electrons); an atomic model proposed by J.J.Thomson in
1904. However, this atomic model had many problems. And it became untenable
as E. Rutherford showed the inconsistencies of such a model in face of his
experiments of alpha particle (ionized helium) scattering by thin sheets of
gold as targets. The main objection to that model being that scattering
experiments indicated a far more ''dilute'' type of matter constituints and
''empty'' space within objects.

Then, Rutherford himself proposed the orbital model for the atom, in which
there was a central nucleus populated by positively charged particles --
protons -- and negatively charge particles -- electrons -- moving around the
nucleus in orbital trajectories, similar to the solar system with the sun at
the center and planets orbiting it, described by the mechanics of celestial
bodies acted on under central forces among them. Later on, the concept of
neutral particles -- neutrons -- within the nucleus came to be added into
the model.

The planetary model, however, was not free of problems. The main objection
was that in such a model, electrons moving around the nucleus would radiate
energy and therefore, classically, such an atom would collapse into itself
after electrons radiated all their energy. Therefore, the model proposed by
Rutherford had to be modified, and here comes the contribution of N. Bohr,
which proposed the hypothesis for quantization of electron orbits making it
possible to better understand the properties of atoms and of the stuff with
which they are made of.

A comment may be in order here: At that time, the atomic nucleus was
understood as a \textquotedblleft storage\textquotedblright\ of mass and
positive charge of the atom. There was no need to know neither a hint as to
what was the internal structure of the nucleus.\cite{01}. These concepts
came into play afterwards, as experiments got more sophisticated and more
deeply concerned as to the internal structure of atoms.

\section{Thomson's model}

In the model proposed by J.J. Thomson \cite{jjt} in 1904, the atom was
considered like a fluid with continuous spherical distribution of positive
charges where electrons with negative charges were embedded, in a number
sufficient to neutralize the positive charges. This model had an implicit
underlying assumption: that of the existence of stable configurations for
the electrons around which they would oscilate. However, according to the
classical electromagnetic theory, there can be no stable configuration in a
system of charged particles if the only interaction among them is of the
electromagnetic character. Moreover, since any electrically charged particle
in an accelerated movement emits electromagnetic radiation, his model had an
additional hypothesis that the normal modes for the oscillating electrons
would have the same frequencies as those observed associated with the lines
of the atomic spectrum. However, there was not found any configuration for
the electrons for any atom whose normal modes had any one of the expected
frequencies. Therefore, Thomson's model for the atom was abandoned because
there was no agreement between its assumptions/predictions and the
experimental results obtained by H. Geiger and E. Marsden \cite{gm}.

\section{The planetary atomic model}

\noindent The discoveries that ocurred by the end of XIX century led the
physicist Ernest Rutherford to do scattering experiments that culminated in
a proposal for the planetary model for atoms.

According to this model, all positive charge of a given atom, with
approximately $99\% $ of its mass, would be concentrated in the atomic
nucleus. Electrons would be moving around the nucleus in circular orbits and
these would be the carriers of the negative charges. Knowing that the charge
of an electron and the charge of a proton are the same in modulus, and that
the nucleus has $Z$ protons, we can define the charge of the nucleus as $%
Ze^{-}$.

Experimentally we observe that in an atom the distance $r$ between the
electron orbit and the nucleus is of the order $10^{-10}m$.

In this section, we build the planetary model for the atom and analyse its
predictions compared to experimental data.

Using Coulomb's law 
\begin{equation}
F=\frac{Z\left\vert e^{-}\right\vert ^{2}}{4\pi \varepsilon _{0}r^{2}}
\label{1}
\end{equation}
and the centripetal force acting on the electron in its circular orbit 
\begin{equation}
F_{c}=\frac{m_{e}v^{2}}{r}  \label{2}
\end{equation}
results in 
\begin{equation}
\frac{Z\left\vert e^{-}\right\vert ^{2}}{4\pi \varepsilon _{0}r^{2}}=\frac{%
m_{e}.v^{2}}{r}  \label{3}
\end{equation}
\begin{equation}
v=\sqrt{\frac{e^{2}}{4\pi \varepsilon _{0}m_{e}r}.Z}  \label{4}
\end{equation}
where we have used the shorthand notation $\left\vert e^{-}\right\vert =e$.

From (\ref{4}) we can estimate the radius for the electron orbit, 
\begin{equation}
r=\frac{e^{2}}{4\pi \varepsilon _{0}m_{e}v^{2}}Z\text{ ,}  \label{4a}
\end{equation}
which means that the radius depends on the total number of protons in the
nucleus, $Z$, and also on the electron's velocity. Here we can make some
definite estimates and see whether our estimates are reasonable, i.e.,
agrees or does not violate experimental data. According to the special
theory of relativity, no greater velocity can any particle possess than the
speed of light. More precisely, for particles with mass like electrons, we
know that their velocity is limited by $v < c$ where $c$ is the speed of
light in vacuum. Substituting for velocity $v=c=3\times10^8 $m/s, electric
charge $e=1.602\times 10^{-19}$m/s, mass of the electron $m_{e}=9.109\times
10^{-31}$kg and $\varepsilon _{0}=8.854\times 10^{-12}$F/m \cite{1a}, we
have 
\[
r>2.813\times 10^{-15}\text{m,} 
\]
where we have taken $Z=1$ and didn't consider the relativistic mass. This
means that we have a lower limit for the radius of an electron's orbit
around the central nucleus, which is consistent with the experimental
observed data where radius of electronic orbits are typically of the order $%
10^{-10}$m.

\subsection{Limitations of the model}

Even though this model could explain some features of the atomic structure
concerning the scattering data, there was nonetheless problems that could
not be explained just by classical mechanics analysis. Since protons and
electrons are charged particles, electromagnetic forces do play their role
in this interaction, and according to Maxwell's equations, an accelerated
electron emits radiation (and therefore energy), so that electrons moving
around the nucleus would be emitting energy. This radiated energy would of
course lead to the downspiraling of electrons around the nucleus until
hitting it. Classical theoretical calculations done predicted that all
electrons orbiting around a nucleus would hit it in less than a second!

However, what we observe is that there is electronic stability, and
therefore the model had to be reviewed.

\section{Bohr's hypothesis}

Analysis of the hydrogen spectrum which showed that only light at certain
definite frequencies and energies were emitted led Niels Bohr to postulate
that the circular orbit of the electron around the nucleus is quantized,
that is, that its angular momentum could only have certain discrete values,
these being integer multiples of a certain basic value \cite{04}. This was
his \emph{\textquotedblleft ad hoc\textquotedblright } assumption,
introduced by hand into the theory. In 1913, therefore, he proposed the
following for the atomic model \cite{Bohr}:

\noindent 1. The atom would be composed of a central nucleus where the
positive charges (protons) are located;

\noindent 2. Around the central nucleus revolved the electrons in equal
number as the positive charges present in the nucleus. The electrons
orbiting such a nucleus had discrete quantized energies, which meant that
not any orbit is allowed but only certain specific ones satisfying the
energy quantization requeriments;

\noindent 3. The allowed orbits also would have quantized or discrete values
for orbital angular momentum, according to the prescription $|\mathbf{L}%
|=n\hbar $ where $\hbar =\displaystyle\frac{h}{2\pi }$ and $n=1,2,3,...$,
which meant the electron's orbit would have specific minimum radius,
corresponding to the angular momentum quantum number $n=1$. That would solve
the problem of collapsing electrons into the nucleus.

Observe that Bohr did not use the value $n=0$ because this does not define
an electronic orbit around the nucleus, although in Plank 's hypothesis

\[
E=nhf 
\]%
where $n=0,1,2,3,...$ the value $n=0$, is perfectly allowed. The reason why
Bohr left out this first quantum number out of his hypothesis comes from
experiment, since spectroscopic studies of many chemical elements show that
these numbers start with $n=1$.

Two colloraries following Bohr's assumptions do follow: First, from item 2.
above, the laws of classical mechanics cannot describe the transition of an
electron from one orbit to another, and second, when electrons do make a
transition from one orbit to another, the energy difference is either
supplied (transition from lower to higher energy orbits) or carried away
(transition from higher to lower energy orbits) by a single quantum of light
- the photon - which has the same energy as the energy difference between
the two orbits.

In this short work we propose that Bohr's atomic orbit quantization
hypothesis is not necessarily needed as an \emph{``ad hoc''} assumption, but
that this can be arrived at using only Planck's assumption of energy
quantization.

First, let us follow the usual pathway where Bohr's quantization is
introduced. Using Newton's second law for the electron moving in a circular
orbit around the nucleus, and thus subjec to Coulomb's law, we have: 
\begin{equation}
\frac{e^{2}}{4\pi \varepsilon _{0}r^{2}}=m\frac{v^{2}}{r}.  \label{c1}
\end{equation}

This allows us to calculate the kinetic energy of the electron in such an
orbit: 
\begin{equation}
E_{\text{k}}=\frac{1}{2}mv^{2}=\frac{e^{2}}{8\pi \varepsilon _{0}r}.
\label{ec1}
\end{equation}

The potential energy for the system proton-electron on the other hand is
given by 
\begin{equation}
E_{\text{p}}=-\frac{e^{2}}{4\pi \varepsilon _{0}r},  \label{ep}
\end{equation}
where $r$ is the radius of the electronic orbit.

Therefore, the total energy for the system is 
\begin{equation}
E=-\frac{e^{2}}{8\pi \varepsilon _{0}r}.  \label{et}
\end{equation}

This result would suggest that, since the radius can have any value, the
same should happen with the angular momentum $L$. 
\begin{equation}
L=pr\sin \theta =pr,\text{ where }\theta =90^{0}  \label{l}
\end{equation}%
that is, the angular momentum depends on the radius. The linear momentum of
the electron is given by 
\begin{equation}
p=mv\text{.}  \label{p}
\end{equation}

Therefore the problem of quantizing the angular momentum $L$ reduces to the
quantizing of the radius $r$, which depends on the total energy (\ref{et}).
Just here Bohr introduced an additional hypothesis, in that the angular
momentum of the electron is quantized, i.e., 
\begin{equation}
L=n\hbar ,  \label{quantl}
\end{equation}%
where $\hbar =\displaystyle\frac{h}{2\pi }$. In this manner he was able to
quantize the other physical quantities such as the total energy. This is the
usual pathway wherein the textbooks normally follow in their sequence of
calculations.

\section{Origin of orbital quantization}

In this work we show another way, in which we do not need the additional
Bohr's assumption input for angular momentum quantization, but stress the
importance of Planck's original energy quantization, from which angular
momentum quantization follows as a consequence. So, from Planck's hypothesis
to quantize the energy (\ref{ec1}) 
\begin{equation}
E=nhf  \label{planck}
\end{equation}%
we have that the total energy for the electron-proton system is quantized
accordingly, i.e., 
\begin{equation}
nhf=E_{\text{n}}  \label{r1}
\end{equation}%
so that if we take the derivative of $E_{\text{n}}$ with respect to the
frequency $f$ we get 
\begin{equation}
\left. \frac{dE_{\text{n}}}{df}\right\vert _{\text{Planck}}=nh.  \label{r2}
\end{equation}

According to the deduction given in the Appendix, modulus its sign, the
total energy for the system electron-proton is

\begin{equation}
E_{\text{n}}=\frac{mv^{2}}{2}  \label{syst}
\end{equation}

Knowing that the scalar orbital velocity is 
\begin{equation}
v=2\pi fr,  \label{r2a}
\end{equation}%
where $f$ is the frequency and $r$ is the radius of the electronic orbit.
Substituting this $v$ into ( \ref{syst} ) we then have:

\begin{equation}
E_{n}=2\pi ^{2}mr^{2}f^{2}  \label{r2b}
\end{equation}%
so the ratio of the system's total energy variation with respect to the
frequency is 
\begin{equation}
\left. \frac{dE_{\text{n}}}{df}\right\vert _{\text{System}}=4\pi ^{2}mr^{2}f.
\label{r3}
\end{equation}%
\qquad \qquad

From (\ref{r2a}), (\ref{l}) and (\ref{p}), we rewrite (\ref{r3}), so that

\begin{equation}
\left. \frac{dE_{\text{n}}}{df}\right\vert _{\text{System}}=2\pi L.
\label{r4}
\end{equation}

Using now (\ref{r2}) in (\ref{r4}) we obtain

\begin{equation}
L=n\hbar.  \label{r5}
\end{equation}

We see that the ratio of the system 's total energy variation with respect
to the frequency plus the Planck's hypothesis of energy quantization leads
to Bohr's assumption.

\section{Conclusions}

In this work we have shown that Planck's fundamental assumption of energy
quantization is more fundamental than Bohr's assumption of angular momentum
quantization. In fact, we have shown that Bohr's rule for angular momentum
quantization can be dispensed of altogether if we consider Planck's energy
quantization and that the former can be derived from this latter one.

The identity (\ref{r2}) can be further clarified by Wilson-Sommerfeld's
rules for quantization \cite{02}. They proposed a set of rules to quantize
any physical system whose coordinates are periodic functions of time. Their
rules are the following: For all physical systems whose coordinates are
periodic functions of time there exists a quantum condition for each
coordinate expressed as: 
\begin{equation}
\oint p_{i}dq_{i}=n_{i}h  \label{ws}
\end{equation}%
where $i$ labels any one of the coordinates, $p_{i}$ the conjugate momentum
associated with such coordinate and $n_{i}$ a quantum number atributed to
such a coordinate.

In our case, we have, 
\[
\oint p_{i}dq_{i}=L\oint d\theta =2\pi L 
\]
and by (\ref{ws}) 
\[
2\pi L=nh 
\]
that is, 
\[
L=n\hbar 
\]
obtaining therefore our result (\ref{r5}).

\section{Appendix}

In this Appendix we will show that the total energy for the system
electron-proton depends on the scalar orbital velocity of the elecron. From
classical mechanics, an electron orbiting a proton in a circular orbit obeys
the folowing equilibrium of forces:

\begin{equation}
\frac{e^{2}}{4\pi \varepsilon r^{2}}=\frac{mv^{2}}{r}  \label{a1}
\end{equation}%
which results in%
\begin{equation}
\frac{e^{2}}{4\pi \varepsilon r}=mv^{2}  \label{a2}
\end{equation}

The total energy $E_{\text{n}}$ for the electron-proton system is equal to
the sum of its kinetic energy $E_{k}$ and its potential energy $E_{p}$.

\begin{equation}
E_{\text{n}}=E_{\text{k}}+E_{\text{p}}  \label{a5}
\end{equation}

where

\begin{equation}
E_{\text{k}}=\frac{mv^{2}}{2}  \label{a6}
\end{equation}

and

\begin{equation}
E_{\text{p}}=-\frac{e^{2}}{4\pi \varepsilon r}  \label{a7}
\end{equation}

The negative sign for the potential energy indicates that our zero reference
for it is at infinity.

Substituting these in (\ref{a5}) we have:

\begin{equation}
E_{\text{n}}=\frac{mv^{2}}{2}-\frac{e^{2}}{4\pi \varepsilon r}  \label{a8}
\end{equation}

From (\ref{a2}), we obtain:

\[
E_{n}=\frac{mv^{2}}{2}-mv^{2}=-\frac{mv^{2}}{2} 
\]

Therefore, the total energy $E_{\text{n}}$ for the electron-proton system is:

\[
E_{\text{n}}=-\frac{mv^{2}}{2} 
\]

Again, the negative sign here defines the bound state of the atom.

\textbf{Acknowledgments:} D. S. Bonaf\'{e} thanks the PIBIC-Jr to the
Universidade Federal de Itajub\'{a}-MG and J.H.O.Sales \ FAPEMIG-CEX 1661/05.

\end{document}